\documentclass[aps,prd,showpacs,eqsecnum]{article}

\usepackage{latexsym}

\begin{document}

\begin{titlepage}
\title{Second Order Post Newtonian Equations of Light Propagation in Multiple Systems}
\author{Chongming Xu\thanks{To whom correspondence should be
addressed, cmxu@pmo.ac.cn; cmxu@njnu.edu.cn} $^{1,2}$, Yanxiang
Gong$^2$, Xuejun Wu$^{1,2}$, Michael
Soffel $^3$ and Sergei Klioner $^3$ \\
$^1$ Purple Mountain Observatory, Nanjing 210008, China \\
$^2$ Department of Physics, Nanjing Normal University, Nanjing
210097, China \\
$^3$ Lohrmann Observatory, Technical University Dresden, D-01062
Dresden, Germany}

\date{\today}

\maketitle

\begin{abstract}
The first order post Newtonian scheme in multiple systems
presented by Damour-Soffel-Xu is extended to the second order one
for light propagation without changing the advantage of the scheme
on the linear  partial differential equations of potential and
vector potential. The spatial components of the metric tensor are
extended to the second order level both in the global coordinates
($q_{ij}/c^4$ term) and in a local coordinates ($Q_{ab}/c^4$
term). The equations of $q_{ij}$ (or $Q_{ab}$) are deduced from
Einstein field equations. The linear relationship between $q_{ij}$
and $Q_{ab}$ are presented also. The 2PN equations of light ray
based on the extended scheme are deduced by means of the iterative
method. We also use parametrized second post Newtonian metric
tensor to substitute into the null geodetic equations  to obtain
the parametrized second order equations of light ray which might
be useful in the observation and measurement in the future space
missions.

\end{abstract}

\quad {\bf Keywords} 2PN Approximate Method, Light Propagation,
Multiple Systems

\end{titlepage}


\setcounter{page}{2}

\section{Introduction}

Recently, in terms of advanced technology a series of space
missions are proposed and planned to launch, e.g. LISA (Laser
Interferometer Space Antenna) \cite{lisa00}, aimed to detect
low-frequency ($10^{-4}$-1 Hz) gravitational wave with a strain
sensitivity of $4 \times 10^{-21}/(Hz)^{1/2}$, GAIA (Global
Astrometric Interferometer for Astrophysics) \cite{perr01} planned
to measure the position and the parallaxes of celestial bodies
with the precision of few $\mu$as (micro arc second), ASTROD
\cite{ni02a} and ASTROD-I \cite{ni02b} (Astro-dynamical Space Test
of Relativity using Optical Devices) in a lower frequency range
and higher sensitivity for the gravitational wave detection
compared to LISA, LATOR (Laser Astrometric Test Of Relativity) to
measure the bending of light near the sun to an accuracy of 0.02
$\mu$as \cite{tury04} and so on. In such a requirement of
experimental precision the second order post Newtonian
contribution to light propagation  has to be taken into account.
On other hand, when the light pass nearby the sun or any planet,
the potential produced by $2^N$ multipole moment is $ V_{\rm mono}
(l / r ) ^N $ \cite{xu97}, where $V _{\rm mono}$ is the potential
created by the mass monopole, $l$ the length deviated to monopole
and $r$ the distance of the action. For the quadrupole of the sun
$(N=2)$ $ ( l / r )^2 < \epsilon ^2 $, where $\epsilon ^2$ is the
small parameter of post Newtonian expansion (nearby the sun
$\epsilon ^2 \sim 10 ^{-6}$), $l$ can be estimated from the
oblateness of the sun \cite{will93}. Therefore if we consider the
second order post Newtonian (2PN) problem, the first post
Newtonian (1PN) quadrupole has to be taken into account. Because
the trajectory of light ray normally is calculated in the global
coordinates, but the relativistic multipole moments have to be
computed in the local coordinates. In 1PN level, the problem has
been thoroughly solved in the scheme presented by Damour-Soffel-Xu
\cite{damo91,damo92,damo93,damo94}, in which a complete 1PN
general relativistic celestial mechanics for N arbitrarily
composed and shaped, rotating deformable bodies is described.
Their scheme is widely abbreviated as the DSX scheme
\cite{reso01}. A similar results has been deduced by means of a
different way: a surface integral derivation \cite{raci04}.

As for light propagation, we should also mention the other
methods. The method of Lorentz covariant theory employed to study
the problems related to the propagation of light ray in a
gravitational field is based on the solution of the null geodesic
equations by means of the first Minkowskian approximation
\cite{kope99,kope02}. In the second order approximation the
Lorentz covariant theory has to be extended also. Numerical
simulation is also an important tool to study the light
propagation in the gravitational field of the moving bodies
\cite{klio03}. A more rigorous approaches (e.g. the second post
Newtonian (2PN) scheme) might be helpful for such kind of
simulations.

The 2PN contribution for light ray has been discussed for a long
time (early 80's) by Epstein and Shapiro \cite{epst80}, Richter
and Matzner \cite{rich82}, and by others (later). But all of them
consider only in one global coordinate system, therefore they can
not calculate the relativistic contribution from multipole moments
which should be calculated in the local coordinates. As we know
the relativistic theory of reference system established only after
1991$^{[10]}$, although it is only 1PN approximation.

In this paper we will extend the 1PN DSX scheme to 2PN for the
discussion on the propagation of light ray, i.e. we will extend
the metric $g_{ij}$ in global coordinates (and $G_{ab}$ in local
coordinates) to $O(6)$ and deduce the corresponding equations from
Einstein field equation. In the section 3, we deduce the second
order post Newtonian equation of light ray by means of the
iterative method. In the section 4, we discuss the parametrized
second post Newtonian formalism (PP$^2$N) which is a tentatively
expansion of Ref.\cite{klio00} and the corresponding equations of
1PN light ray. Some conclusion remark is made in Sec. 5.


\section{The extension of DSX scheme}

In this section, we will extend 1PN DSX scheme to 2PN scheme for
discussion on the second order contribution in light propagation.
Our symbols and signature follow the DSX scheme. Here we summarize
the notation in DSX paper which we will  use in this paper. The
signature -+++ is taken; spacetime indices go from 0 to 3 and
denoted by Greek indices, while space indices (1 to 3) are denoted
by Latin indices. We use Einstein's summation convection for both
types of indices, whatever the position of repeated indices. The
flat metric is denoted by $f_{\mu\nu}$, with components diag
(-1,+1.+1,+1) in Lorentzian coordinates. In post Newtonian
expansions we shall often abbreviate the order symbol $O(c^{-n})$
simply by $O(n)$. The ``global" ( or ``common view") coordinates
used for describing the overall dynamics of the system will be
denoted by $(x^\mu)\equiv (ct, x^i)$. By contrast, each of the
``local" coordinate systems $A$, used for describing the internal
dynamics of each body, will be denoted by $(X^\alpha_A)\equiv
(cT_A, x^a_A)$. We distinguish the second part of the Latin
alphabet (i,j,k, $\cdots$) for global space coordinates from the
first part of the Latin alphabet (a,b,c, $\cdots$) for local space
coordinates as done in the DSX scheme. A spatial multi-index
containing $l$ indices is simply denoted $L$ (and $K$ for $k$
indices, etc.), i.e., $L \equiv i_1 i_2 \dots l_l$. A
multi-summation is always understood for repeated multi-index
$S_LT_L = \Sigma_{i_1 \dots l_l} S_{i_1 \dots i_l}T_{i_1 \dots
i_l}$. Given a spatial vector, $v^i$, its $l \rm{th}$ tensorial
power is denoted by $v^L \equiv v^{i_1} v^{i_2} \dots v^{i_l}$.
Also, $\partial _L \equiv \partial _{i_1} \dots
\partial _{i_l}$. The symmetric and trace-free (STF) part of a
spatial tensor will be denoted by angular brackets (or by a caret
when no ambiguity arises): $STF_{i_1 \dots i_l}(T_{i_1 \dots i_l})
\equiv T_{(i_1 \dots i_l)} \equiv \stackrel{\wedge}{T}_L$. The
coordinate transformation between the local coordinate system
$X^\alpha_A$ and the global one $x^\mu$ reads (omitting a
labelling index $A$ on all quantities pertaining to the local
frame):
\[
x^\mu = f^\mu (X^a)=z^\mu(X^0) + e^\mu_a (X^0) Y^a(X^0,X^b)
  + \xi ^\mu
\]
where
\[
Y^a(X^0,X^b) \equiv X^a + (1/c^2) [ (1/2)A^a(X^2) - X^a(A_bX^b)].
\]
Moreover, one has the definitions %
\begin{equation} \label{1eq}
A_a(S)=f_{\mu\nu} e^\mu_a (S) \frac{d^2 z^\nu}{d\tau ^2} \, ,
\quad v^i= c\frac{dz^i}{dz^0}\, , \quad e^i_0 = c^{-1}e^0_0 v^i \,
, \quad e^0_a (S) = e^i_a \frac{dz^i}{dS} + O(4)
\end{equation}
and
\begin{eqnarray}
e^0_0 (S) &=& 1 + \frac{1}{c^2}\left( \frac{1}{2} v^2 +
\overline{w}
  \right) + \frac{1}{c^4} \left[ \frac{3}{8} v^4
  + \frac{1}{2} (\overline{w})^2
  + \frac{5}{2}\overline{w} v^2 - 4 \overline{w_i} v^i \right]
  + O(6) \, , \label{2eq} \\
e^0_a (S) &=& R^i_a \left\{ \frac{v^i}{c} \left[ 1 +
\frac{1}{c^2}\left( \frac{1}{2} v^2 + 3\overline{w} \right)
\right]
  - \frac{4}{c^4} \overline{w_i} \right\}
  + O(5) \, , \label{3eq} \\
e^i_a (S) &=& \left[ 1 - \frac{1}{c^2} \overline{w} \right] \left[
  \delta ^{ij} + \frac{1}{2c^2} v^iv^j \right]  R^i_a
  + O(4) \, , \label{4eq}
\end{eqnarray}
where $R^i_a(S)$ is a slowly changing rotation matrix $R^i_aR^j_a
= \delta ^{ij}$, $R^i_aR^i_b = \delta ^{ab}$, $c dR^i_a /dS =
O(2)$. All of above three equations are already shown in
Eq.(5.21b)--(5.21d) of Ref.\cite{damo91}, as for Eq.(\ref{4eq})
will be extended to $O(6)$ later (see Eq.(\ref{51eq})). Now, in
this paper, a fixed-star coordinates has been chosen, therefore
$R^i_a = \delta ^i_a$, $z^\mu$ represents each central world line
${\mathcal{L}}_A (X^a = 0)$. $\overline{w}$ is the external
potential of body A, $v^i$ is the coordinate three-velocity of the
central world line ${\mathcal{L}}_A$ measured in the global
coordinate system, and $V^a$ defined by $v^i=R^i_a V^a$ or
$V^a=R^a_i v^i$.

As we mentioned early, DSX scheme is the first post Newtonian
approximation for particle motion. The metric tensor in a global
coordinates is written in the form
\begin{eqnarray}
g_{00} &=& - \exp  \left( - \frac{2w}{c^2} \right) + O(6) \, ,
 \label{5}\\
g_{0i} &=& -  \frac{4w_i}{c^3} + O(5) \, , \label{6}\\
g_{ij} &=& \delta _{ij} \exp \left( \frac{2w}{c^2} \right)
 + O(4) \, ,  \label{7}
\end{eqnarray}
and they satisfy the conformal isotropic condition
\begin{equation}\label{8}
 g_{00} g_{ij} = - \delta _{ij} + O(4) \, .
\end{equation}
The metric tensor in the local coordinates A has a similar form as
in global, one only needs to change the small letters into capital
letters
\begin{eqnarray}
G_{00} &=& - \exp \left( - \frac{2W}{c^2} \right) + O(6) \, ,
  \label{9}\\
G_{0a} &=& - \frac{4W_a}{c^3} + O(5) \, , \label{10}  \\
G_{ab} &=& \delta _{ab} \exp \left( \frac{2W}{c^2} \right)
         + O(4) \, , \label{11}
\end{eqnarray}
and they also satisfy the conformal isotropic condition
\begin{equation}\label{12}
 G_{00} G_{ab} = - \delta _{ab} + O(4) \, .
\end{equation}
The potential $W$ and vector potential $W_a$ for body A can be
divided into two parts (self part $W^+, W^+_a$ and external part
$\overline{W}$, $ \overline{W}_a$)
\begin{eqnarray}
W &=& W^+ + \overline{W} \, , \label{13} \\
W_a &=& W^+_a + \overline{W}_a \, . \label{14}
\end{eqnarray}
As in DSX scheme, we use $W_\alpha$ to represent $ W$ ($\alpha=0$)
and $W_a$ ($\alpha = a$). From Einstein field equation, we obtain
equations to be satisfied by $ W$ and $W_a$ (compare Eq.(4.3a),
(4.3b) and (4.4) of Ref.\cite{damo91})
\begin{eqnarray}
&& \Box _X W + \frac{4}{c^2} \partial _T (\partial _T W
  + \partial _b W_b) =  - 4 \pi G \Sigma  + O(4) \, , \label{15}
  \\
&& \Delta _X W_a - \partial _a (\partial _T W
  + \partial _b W_b) = - 4 \pi G \Sigma ^a + O(2) \, , \label{16}
\end{eqnarray}
where $\Box _X = \Delta  _X - c^{-2} \partial ^2 _T$ with $ \Delta
_X = \partial ^2 / \partial X^a \partial X^a$, and where the
source terms
\begin{equation} \label{17eq}
\Sigma ^a \equiv (\Sigma , \Sigma ^a) \equiv \left| \frac{T^{00}
  + T^{aa}}{c^2}, \frac{T^{0a}}{c} \right|
\end{equation}
are now defined by components of the stress-energy tensor in the
$X^a$ coordinate system.

In harmonic gauge
\begin{equation}\label{18eq}
\partial _T W + \partial _a W_a = 0 \, .
\end{equation}
We have (see Eq.(4.51) of Ref.\cite{damo91})
\begin{equation}\label{19eq}
\Box _X W^+_\alpha = - 4\pi G \Sigma _\alpha \, .
\end{equation}
By means of Blanchet-Damour multipole moments \cite{blan89} we
obtain the solutions
\begin{eqnarray}
W^{+A}(T,{\bf X}) &=& G \sum_{l\geq 0} \frac{(-1)^l}{l!}
  \partial_L [ R^{-1}M^A_L (T\pm R/c)]
  + \frac{1}{c^2} \partial _T \Lambda^A
  + O(4) \, , \label{20eq} \\
W^{+A}_a(T,{\bf X}) &=& - G \sum_{l\geq 1} \frac{(-1)^l}{l!}
  \left[ \partial _{L-1}
  \left( R^{-1} \frac{d}{d T}M^A_{aL-1} (T\pm R/c)\right)
  + \frac{l}{l+1} \epsilon_{abc}\partial _{bL-1}
  [ R^{-1}S^A_{cL-1}(T\pm R/c)] \right] \nonumber \\
  &&  - \frac{1}{4} \partial_a\Lambda^A
  + O(2) \, , \label{21eq}
\end{eqnarray}
where
\begin{eqnarray}
&& \Lambda^A \equiv 4\pi \sum_{l\geq 0} \frac{(-1)^l}{(l+1)!}
  \frac{2l+1}{2l+3} \partial_L [R^{-1}\mu^A_L(T\pm R/c)]
  \, , \label{22eq} \\
&& \mu^A_L(T)\equiv \int_A d^3X\stackrel{\wedge}{X}{}^{bL}
  \Sigma^b(T,{\bf X}) \, . \label{23eq}
\end{eqnarray}
In Eqs.(\ref{20eq})--(\ref{23eq}) the $\pm$ sign in a function of
the local time, $T$, divided by the coordinate distance to the
origin of the local system, $R\equiv |{\bf X}|$, denotes the
average
\begin{equation} \label{24eq}
\frac{F(T \pm R/c)}{R} \equiv \frac{1}{2} \left(
  \frac{F(T-R/c)}{R} + \frac{F(T+R/c)}{R} \right) \, ,
\end{equation}
which is the well-known time-symmetric solution, with spherical
symmetry, of the wave equation.

The ``mass" ($M^A_L$) and ``spin" ($S^A_L$) multipole moments of
body A appearing in Eqs.(\ref{20eq}), (\ref{21eq}) are the STF
Cartesian tensors defined by the same expressions of the matter
distribution variables \cite{blan89} for closed gravitationally
self-interacting system, but now restricted to an integration over
the volume of body A, using the local-system matter variables
Eq.(\ref{17eq}): i.e.,
\begin{eqnarray}
M^A_L(T) &\equiv & \int_A d^3X\stackrel{\wedge}{X}^L
  \Sigma (T,{\bf X}) + \frac{1}{2(2l+3)c^2}
  \frac{d^2}{dT^2} \left(
  \int_A d^3X\stackrel{\wedge}{X}{}^L {\bf X}^2
  \Sigma (T,{\bf X}) \right) \nonumber \\
  &&  - \frac{4(2l+1)}{(l+1)(2l+3)c^2}  \frac{d}{dT} \left(
  \int_A d^3X\stackrel{\wedge}{X}{}^{aL}
  \Sigma ^a (T,{\bf X}) \right) + O(4)\, (l\geq 0)
  \, , \label{25eq} \\
S^A_L(T) &\equiv & \int_A d^3X \epsilon^{ab<c_l}
  \stackrel{\wedge}{X}{}^{L-1>a}\Sigma^b(T,{\bf X})
  + O(2) \, (l\geq 1) \, . \label{26eq}
\end{eqnarray}
In other hand, from the definition of the external part in the
global coordinates, we have in the global coordinates for N bodies
system
\begin{equation} \label{27eq}
\overline{w}^A_\mu= \sum^N_{B\neq A} w^B_\mu \, ,
\end{equation}
where $w^B_\mu$ is the four potentials of the body B . From now on
we will omit the labelling index $A$ on all quantities pertaining
to the local frame. Based on $\overline{w}_\mu$, we can calculate
$\overline{W}_\alpha $ with the following equations
\begin{eqnarray}
&& w=\left( 1 + \frac{2V^2}{c^2} \right) W
  + \frac{4}{c^2} V^a W_a
  + \frac{c^2}{2} \ln ( A^0_0 A^0_0
  - A^0_a A^0_a ) + O(4) \, , \label{28eq} \\
&& w _i = v^i W + R^i_a W_a
  + \frac{c^3}{4} ( A^0_0 A^i_0
  - A^0_a A^i_a ) + O(2) \, . \label{29eq}
\end{eqnarray}
Then from some calculation we can get $\overline{W}$ and $
\overline{W}_a$ also. Up to now we have briefly reviewed the main
story of DSX scheme. But as we mentioned early that DSX scheme is
1PN approximation method (for the equations of motion of bodies).
If we want to extend DSX scheme to apply to the second order post
Newtonian (2PN) approximation of light propagation, we have to
extend the metric $g_{ij}$ to $O(6)$ level rather than $O(4)$
\cite{xu03}, i.e.
\begin{eqnarray}
G_{00} &=& - \exp \left( - \frac{2W}{c^2} \right) + O(6) \, ,
  \label{30eq}\\
G_{0a} &=& - \frac{4W_a}{c^3} + O(5) \, , \label{31eq}  \\
G_{ab} &=& \delta _{ab} \exp \left( \frac{2W}{c^2} \right)
         + \frac{Q_{ab}}{c^4}+ O(6) \, . \label{32eq}
\end{eqnarray}
The contravariant metric tensor reads
\begin{eqnarray}
G^{00} &=& - \exp \left(  \frac{2W}{c^2} \right) + O(6) \, ,
   \label{33eq}\\
G^{0a} &=& - \frac{4W_a}{c^3} + O(5) \, , \label{34eq}  \\
G^{ab} &=& \delta _{ab} \exp \left( - \frac{2W}{c^2} \right)
         - \frac{Q_{ab}}{c^4}+ O(6) \, .  \label{35}
\end{eqnarray}
The metric tensor in the global coordinate system has a similar
form, just change $G_{\alpha\beta}$, $W$, $W_a$ and $Q_{ab}$ by
$g_{\mu\nu}$, $w$, $w_i$ and $q_{ij}$.

The relations between $w \, , \; w_i \, , \; q_{ij} $ and $W \, ,
\; W_a \, , \; Q_{ab} $ will be studied in the following. In DSX
scheme, since the metric in the global coordinate system as well
as in every local coordinate systems has the similar form, so
that, if we deduce any equation in one local coordinate system,
then we have it in every local coordinates as well as in the
global coordinates. Now in this paper, the situation is similar,
Therefore we need only to treat in a local coordinate system A,
but we omit the index A always. The spatial conformal isotropic
condition Eq.(\ref{12}) is revised as
\begin{equation}\label{36eq}
G_{ab} G_{00} = - \delta _{ab} - \frac{Q_{ab}}{c^4} + O(6) \, .
\end{equation}
If we attribute $Q_{ab}/c^4$ to $O(4)$, it is just the form in DSX
scheme. Therefore  Eq.(\ref{32eq}) is an extension of
Eq.(\ref{7}). $Q_{ab}$ also can be taken as a spatial anisotropic
contribution in second order (see Eq.(\ref{36eq})).

The metric tensor $G_{0a}$ is taken as $O(3) $ here, in fact it is
only as small as $O(4)$ because of the relatively slow rotation
rate in the sun \cite{rich82,rich81}. To keep the uniformity with
DSX scheme therefore we take such formula of $G_{0a}$ as well as
for $g_{0i}$.

We have to calculate the equations satisfied by $W$, $W_a$ and
$Q_{ab}$ from the Einstein field equations. On the first step, the
Christoffel symbols can be calculated from the metric tensor
\begin{eqnarray}
\Gamma ^0_{00} &=& - \frac{W_{,t}}{c^3} + O(5) \, , \nonumber\\
\Gamma ^0_{0a} &=& - \frac{W_{,a}}{c^2} + O(6) \, , \nonumber\\
\Gamma ^a_{00} &=& - \frac{W_{,a}}{c^2}
  + \frac{4 WW_{,a}}{c^4} - \frac{4W_{a,t}}{c^4}+ O(6) \, , \nonumber\\
\Gamma ^0_{ab} &=& \delta _{ab} \frac{W_{,t}}{c^3}
  + \frac{4}{c^3} W_{(a,b)} + O(5) \, , \nonumber\\
\Gamma ^a_{0b} &=& - \frac{4}{c^3} W_{[a,b]}
  + \frac{W_{,t}}{c^3} \delta _{ab} + O(5) \, , \nonumber\\
\Gamma ^a_{bc} &=&  \frac{1}{c^2} (\delta _{ab} W_{,c}
  + \delta _{ac} W_{,b} - \delta _{bc}W_{,a} )
  +  \frac{1}{2c^4} ( Q_{ab,c}
  + Q_{ac,b} - Q_{bc,a} ) + O(6) \, ,\label{37eq}
\end{eqnarray}
where two indices enclosed in a parentheses is denoted as
symmetrization, and in a square bracket means antisymmetrization.

Then we can deduce the Ricci tensor
\begin{eqnarray}
R^{00} &=& - \frac{\nabla ^2 W}{c^2} - \frac{1}{c^4}
  \left( 3\partial _t \partial _t W
  + 4\partial _t \partial _d W_d \right) + O(6) \, ,  \label{38eq} \\
R ^{0a} &=& - \frac{2}{c^3} \left( \nabla ^2 W _a
  - \partial _a \partial _d W_d
  - \partial _t \partial _a W \right) + O(5) \, , \label{39eq}\\
R ^{ab} &=& - \frac{1}{c^2} \delta _{ab} \nabla ^2 W
  + \frac{1}{c^4} \left[ 4 \delta _{ab} W \nabla ^2 W
  +  \delta _{ab} W_{,tt}
  + 4 W_{(a,b),t} - 2W_{,a}W_{,b} \right. \nonumber \\
  && \left. + \frac{1}{2} ( Q_{ad,bd} + Q_{bd,ad}
  - Q_{ab,dd} - Q_{dd,ab}) \right] + O(6) \, , \label{40eq}
\end{eqnarray}
and the scalar curvature
\begin{equation}
R = - \frac{2}{c^2} \nabla ^2 W + \frac{1}{c^4}
  ( 4W \nabla ^2 W + 6 W_{,tt} + 8 W_{a,at}
  - 2 W_{,a} W_{,a} + Q_{ab,ab} - Q_{aa,bb} ) +O(6) \, . \label{41eq}
\end{equation}

From Einstein field equation we can derive equations of $W$, $W_a$
and $Q_{ab}$
\begin{eqnarray}
&& \nabla ^2 W + {1 \over c^2} ( 3W_{,tt}
  + 4 \partial _t \partial _a W_a) = - 4 \pi G \Sigma
  + O(4) \, , \label{42eq}\\
&& \nabla ^2 W_a - \partial _a \partial _b W_b
  -  \partial _t \partial _a W = - 4 \pi G \Sigma ^a
  + O(2) \, , \label{43eq} \\
&& -2 \delta _{ab} W _{,tt} + 4( W_{(a,b),t}
  - \delta _{ab} W _{(d,d),t} ) -2 W_{,a} W_{,b}
  + \delta _{ab} W_{,d}W_{,d} \nonumber \\
&& \quad + \frac{1}{2} ( Q_{ad,bd} + Q_{bd,ad}
  - Q_{ab,dd} - Q_{dd,ab} - \delta _{ab} Q_{dc,dc}
  + \delta _{ab} Q_{dd,cc} ) = 8 \pi G T ^{ab} \, , \label{44eq}
\end{eqnarray}
where $ \Sigma = ( T^{00} + T ^{aa} ) / c^2 $ and $ \Sigma ^a =
T^{0a} /c $ (the same definition in DSX scheme).
Eq.(\ref{42eq}) and (\ref{43eq}) are linear PDE, which can be
solved in certain suitable gauge conditions, then Eq.(\ref{44eq})
will be solved also. We shall point out that, $T^{ab}$ can be
expressed by $\Sigma$ and $\Sigma ^a$ (to see Eq.(A13) of
Ref.\cite{xu01}). Therefore it is self-consistent within the
framework of DSX scheme. Similar field equations in the global
coordinate system can be obtained easily by substituting capital
letters by small letters.

From Eqs.(\ref{42eq}) and (\ref{43eq}) we see that $Q_{ab}$ does
not appear in the field equations of $W$ and $W_a$, and they are
the same as ones in DSX scheme (keeping linear equations). The
solutions of $W$ and $W_a$ related to relativistic multipole
moments are still valid as before in DSX scheme.

Now, we shall discuss the transformation relations of the
potential, the vector potential and $Q_{ab}$ (or $q_{ij}$) from
the global coordinate system to a local coordinate system and vice
versa. The coordinate transformation law reads
\begin{equation}\label{45eq}
g^{\mu\nu} = \frac{\partial x^\mu}{\partial X^\alpha}
  \frac{\partial x^\nu}{\partial X^\beta} G ^{\alpha\beta} \, .
\end{equation}
Considering $g^{00}$, we obtain
\begin{equation}\label{46eq}
w=\left( 1 + \frac{2V^2}{c^2} \right) W
  + \frac{4}{c^2} V^a W_a
  + \frac{c^2}{2} \ln ( A^0_0 A^0_0
  - A^0_a A^0_a ) + O(4) \, ,
\end{equation}
where $A^\mu_{\alpha} = \partial x ^\mu / \partial X ^\alpha$, $ V
^a= R ^a _i  v^i$ ($R^a _i$ is a slowly changing rotation matrix
and $v^i $ is the coordinate three-velocity of the central world
line
measured in the global coordinate system). \\
For $g^{0i}$, it turns out to be
\begin{equation}\label{47eq}
w _i = v^i W + R^i_a W_a
  + \frac{c^3}{4} ( A^0_0 A^i_0
  - A^0_a A^i_a ) +O(2) \, .
\end{equation}

In Eqs.(\ref{46eq}) and (\ref{47eq}), $Q_{ij}$ does not enter
these equations, and the relations between $w$, $w_i$ and $W$,
$W_a$ are the same as in DSX scheme (to keep linear relation). As
we know, all of the discussion about the theory of reference
systems are based on the linear PDE of potentials
(Eq.(\ref{42eq}), (\ref{43eq})) and linear relation between
potentials in the global coordinates and in the local coordinates
((\ref{46eq}) and (\ref{47eq})). Since these equations are the
same as before (in DSX scheme), the theory of reference system in
DSX scheme (for potential and vector potential) is valid in this
paper. Also in Eqs.(\ref{42eq}), (\ref{43eq}), (\ref{46eq}) and
(\ref{47eq}) $Q_{ab}$ does not appear, therefore self-parts $W^+$,
$W^+_a$ and external parts  $\overline{W}$, $\overline{W}_a$ can
be obtained just like in the DSX scheme \cite{damo91,damo92}.

Finally, for $g^{ij}$, we get
\begin{eqnarray}
q _{ij} &=& -2W ( 2 V^2 \delta _{ij}  - R^i_a  R^j_b
  V^a V^b ) - 8 V^a W _a \delta _{ij}
  + 8 R_a ^{(i} R ^{j)} _b W ^a V ^b
  + R^i_a R^j_b Q_{ab} \nonumber \\
  && + 2 W c^2 (A^i _a A ^j _ a - \delta _{ij})
  + c^4 \{  A^i_0 A^j_0 - A^i_a A^j_a
  + \delta _{ij} [ 1 - \ln ( A^0_0 A^0_0
  - A^0_a A^0_a )  ] \} +O(2) \, . \label{48eq}
\end{eqnarray}
Here we should emphasize that Eq.(\ref{48eq}) has only formally
given for completeness.

We should point that in Ref.\cite{xu03}, our calculation is
incomplete since in Eq.(\ref{48eq}) $A^j_a$ must be calculated up
to $O(6)$ level, but in fact $A^j_a$ is related with $e^j_a$ as
\begin{equation} \label{49eq}
A^j_a = e^j_a + \frac{\partial \xi ^j}{\partial x^a} \, ,
\end{equation}
where $e^j_a$ is the value of $A^j_a$ at the central world line of
body A. But precision of $e^j_a$ is only $O(4)$ level (see
Eq.(\ref{4eq})), therefore we need to extend $e^j_a$ to $O(6)$. We
have the formula (see Eq.(5.19) of Ref.\cite{damo91})
\begin{equation} \label{50eq}
\overline{g}_{\alpha\beta}e^\alpha _a e^\beta _b
  = \delta _{ab} \,.
\end{equation}
From Eq.(\ref{50eq}) and (\ref{4eq}), we have
\begin{equation} \label{51eq}
e^i_a = R^i_a - \frac{R^i_a}{c^2} \overline{w} |_A +
\frac{1}{2c^2}
  v^i_A V_a + \frac{b^i_a}{c^4} + O(6) \, ,
\end{equation}
where
\begin{equation} \label{52eq}
b^i_a =\frac{1}{2} \left. \overline{w}^2 \right| _A R^i_a
  + \frac{3}{8} V^2 V_a v^i
  + \frac{3}{2} \left. \overline{w} \right| _A V_a v^i
  - \frac{1}{2} \left. \overline{q}_{ij}\right|_A R^j_a \, .
\end{equation}
where $\left. \right|_A$ denote the value at the central world
line of body A. In Eq.(\ref{44eq}), $W$ and $W_a$ can be taken as
known functions. By means of harmonic gauge ($\partial _a W_a +
\partial _t W =0$),
Eq.(\ref{44eq}) becomes %
\begin{eqnarray}
&& 2 \delta _{ab} W _{,tt} + 4 W_{(a,b),t}
   -2 W_{,a} W_{,b} + \delta _{ab} W _{,d} W_{,d}
   \nonumber \\
&& \quad + \frac{1}{2} \left( Q_{ad,bd} + Q_{bd,ad}
  + \delta _{ab} Q_{dd,cc}
  - Q_{ab,dd} - Q_{dd,ab} - \delta _{ab} Q_{dc,dc}
  \right) = 8 \pi G T ^{ab} \, ,. \label{53eq}
\end{eqnarray}
We also divide $Q_{ab}$ into self-part and
external part:
\begin{eqnarray} \label{54eq}
Q_{ab} = Q^+_{ab} + \overline{Q}_{ab} \, .
\end{eqnarray}
In DSX scheme, they took the ``weak effacement of post Newtonian
external gravitational potentials in the local frame", i.e. the
``external" PN potentials $\overline{W}^a_\alpha (T_A,X_A)$ in the
local A frame vanish for all $T_A$ times, at the origin of the
frame (at the central world line):
\begin{equation} \label{55eq}
\forall \; T_A, \quad  \overline{W}^A_\alpha (T_A, 0, 0, 0) =0
\end{equation}
(see Eq.(5.12) of Ref.\cite{damo91}). We assume ``weak effacement"
condition to be valid also for $\overline{Q}_{ab}$, i.e.
\begin{equation} \label{56eq}
\forall \; T_A, \quad \overline{Q}^A_{ab} (T_A, 0, 0, 0) =0 \, .
\end{equation}
With the weak effacement condition, we directly write out
equations satisfied by $Q^+_{ab}$ and $\overline{Q}_{ab}$ from
Eq.(\ref{53eq}):
\begin{eqnarray}
&& 2 \delta _{ab} W^+ _{,tt} + 4 W^+_{(a,b),t}
   -2 W^+_{,a} W^+_{,b} -2 ( W^+_{,a} \overline{W}_{,b}
   + \overline{W}_{,a} W^+_{,b}) + \delta _{ab} W^+_{,d} W^+_{,d}
   + 2\delta_{ab}W^+_{,d} \overline{W}_{,d}  \nonumber \\
&& \qquad \qquad + \frac{1}{2} \left( Q^+_{ad,bd} + Q^+_{bd,ad}
  + \delta _{ab} Q^+_{dd,cc}  - Q^+_{ab,dd}
  - Q^+_{dd,ab} - \delta _{ab} Q^+_{dc,dc}
  \right) = 8 \pi G T ^{ab} \, , \label{57eq}\\
&& 2 \delta _{ab} \overline{W} _{,tt} + 4 \overline{W}_{(a,b),t}
   -2 \overline{W}_{,a} \overline{W}_{,b}
   + \delta _{ab} \overline{W} _{,d} \overline{W}_{,d}
   \nonumber \\
&& \qquad \qquad + \frac{1}{2} \left( \overline{Q}_{ad,bd}
  + \overline{Q}_{bd,ad} + \delta _{ab} \overline{Q}_{dd,cc}
  - \overline{Q}_{ab,dd} - \overline{Q}_{dd,ab}
  - \delta _{ab} \overline{Q}_{dc,dc}
  \right) = 0  \, . \label{58eq}
\end{eqnarray}
The weak effacement condition is automatically satisfied for
Eq.(\ref{58eq}). In principle from Eqs.(\ref{57eq}) and
(\ref{58eq}) and boundary condition we could solve $ Q^+_{ab}$ and
$\overline{Q}_{ab}$ numerically. But for the problem of the light
propagation in the solar system it becomes much simple. Since the
velocity of the relative motion is small inside the sun ($v^2 /
c^2 < \epsilon ^2$), especially the shape of the sun is close to a
monopole, therefore if the origin of the coordinate system is
taken as the center of the solar mass (dipole always equal to
zero), the quadrupole terms (and higher multipole moments) in 2PN
contribution on $Q_{ab}$ can be ignored, i.e.
\begin{equation}\label{59eq}
Q_{ab} = \delta _{ab} Q/3
\end{equation}
in the local coordinates of the sun. As for other local
coordinates of each planets 2PN contribution is too small
($10^{-19}-10^{-18}$) to be calculated. In  the local coordinates
of the sun the external potential is negligible, then $W=W^+$.

Substituting Eq.(\ref{59eq}) into Eq.(\ref{53eq}) and taking PN
gauge ($ 3\partial _t W + 4 \partial _a W_a = O(2)$), $Q$
satisfies the equation as following:
\begin{equation}\label{60eq}
\left( \frac{Q}{3} + {1 \over 2} W^2 \right) _{,dd} = 8\pi G
\left( T^{aa}
  - \frac{W \Sigma}{2} \right) \, .
\end{equation}
The right hand side of above equation has a compact support source
and the solution of Eq.(\ref{60eq}) is
\begin{equation}\label{61eq}
Q = -  \int \frac{6G ( T^{aa}- W \Sigma /2)}{r} d^3 X
  - {3 \over 2} W ^2 \, .
\end{equation}
From  Eq.(\ref{61eq}) it is clear that $Q$ itself does not have
compact support neither is linear in $W$ (because of the last
term). We should point that if we use PN gauge, $W$ and $W_a$
still can use B-D moments to expand , but one needs to do the
gauge transformation from the harmonic gauge to the PN gauge by
means of choosing a suitable $\Lambda $ in Eqs.(\ref{20eq}) and
(\ref{21eq}).

In the global coordinates of the solar system, we have a similar
equation and solution. Since there is very small difference
between the global coordinates of the solar system and the local
coordinates of the sun (additional $10^{-3}$ in $c^{-4}$ level) we
can always ignore the distinction between $q$ and $Q$ of the sun.
Therefore, for the situation of the solar system, it may be
unnecessary to solve Eqs.(\ref{57eq}) and (\ref{58eq}) directly,
but in the following discussion we will still keep $q_{ij}$ in the
equations for the generality.


\section{2PN Equations of Light Ray}

In this section we will deduce the 2PN equations of light ray in
the global coordinates by the iterative method. Certainly, the
result can be also used in the local coordinates by substituting
the quantities in the global coordinates for the corresponding
quantities in the local coordinates. For example, by replacing
$w$, $w_i$, $q_{ij}$, $g_{\mu\nu}$ by $W$, $W_a$, $Q_{ab}$,
$G_{ab}$. We start the basic equations of light ray: %
\begin{eqnarray}
&&\quad  g_{\mu\nu} \frac{dx^\mu}{d\lambda}
  \frac{dx^\nu}{d\lambda } = 0 \, , \label{3.1} \\
&& \frac{d^2x^\mu}{d\lambda ^2} + \Gamma _{\nu\sigma }^\mu
  \frac{dx^\nu}{d\lambda }\frac{dx^\sigma }{d\lambda } = 0
  \, ,  \label{3.2}
\end{eqnarray}
where $\lambda$ is an ``affine" parameter. Normally we replace
$\lambda$ by time $t$ in terms of $\mu =0$ component of
Eq.(\ref{3.2}) (see Ref.\cite{will93}), then Eq.(\ref{3.1}) and
(\ref{3.2}) become
\begin{equation} \label{3.3}
\frac{d^2x^i}{dt ^2} = \left( \frac{1}{c} \Gamma^0 _{\nu\sigma }
  \frac{dx^i}{dt } - \Gamma ^i_{\nu\sigma} \right)
  \frac{dx^\nu}{dt} \frac{dx^\sigma }{dt}  \, ,
\end{equation}
and
\begin{equation} \label{3.4}
g_{\mu\nu} \frac{dx^\mu}{dt}\frac{dx^\nu}{dt} = 0 \, .
\end{equation}
In fact Eqs.(\ref{3.3}) and (\ref{3.4}) are our real basic
equations of light ray. For 1PN equation of light ray, we use 1PN
metric (neglect $g_{0i}$ and all $O(4)$ terms in $g_{ij}$ and
$g_{00}$) and corresponding Christoffel symbols Eq.(\ref{37eq}):
\begin{eqnarray}
 g_{00} &=& - 1 + \frac{2\omega }{c^2} + o(4) \, ,
   \label{3.5a} \\
 g_{ij} &=& \delta _{ij} \left( 1 + \frac{2\omega }{c^2} \right)
   + o(4) \, . \label{3.5b}
\end{eqnarray}
Then Eqs.(\ref{3.3}) and (\ref{3.4}) become %
\begin{eqnarray}
&& c^2 - 2w - \left( 1 + \frac{2w}{c^2} \right)
  \left| {\frac{d{\bf x}}{dt}} \right|^2=0 \, ,\label{3.6a}\\
&& \frac{d^2x^i}{dt^2} = w _{,i} \left( 1 + \frac{1}{c^2}
  \left| {\frac{d{\bf x}}{dt}} \right|^2 \right)
  - \frac{4}{c^2}\frac{dx^i}{dt} \left( \frac{d{\bf x}}{dt}
  \cdot \nabla w \right) \, . \label{3.6b}
\end{eqnarray}
The solution of equations in 1PN level is
\begin{equation} \label{3.7}
{\bf x} = {\bf x}_0 + c {\bf n} (t - t_0) + {\bf x}_{1P}\, ,
\end{equation}
where ${\bf x}_{1p}$ is 1PN revision based on the Newtonian or the
zeroth order solution $({\bf x} = {\bf x}_0 + c {\bf n} (t - t_0)
)$ and ${\bf n}$ is the directional unit vector which is given in
the initial conditions. Finally, we have
\begin{eqnarray}
&& c {\bf n} \cdot {\frac{d{\bf x}_{1P}}{dt}}=-2w \, ,\label{3.8a}\\
&& \frac{d^2{\bf x}_{1P}}{dt^2} = 2 \left[ \nabla w
  - 2 {\bf n} \left( {\bf n} \cdot \nabla w \right) \right]
  \, . \label{3.8b}
\end{eqnarray}
Eqs.(\ref{3.8a}) and (\ref{3.8b}) agree to the parametrized 1PN
equations of light ray deduced by will \cite{will93}, when we put
$\gamma =1 $ and replace Will's $U$ by $w$ in DSX scheme. Based on
1PN solution of light ray (${\bf x}_{1P}$ as a known function), we
deduce the 2PN equations of light ray in the extended DSX scheme
by using the iterative method. Substituting the 2PN metric tensor
in global coordinates (Eqs.(\ref{30eq}-\ref{32eq}) replaced
$G_{\alpha\beta}, \, W, \, W_a$ and $Q_{ab}$ by $g_{\mu\nu}, \, w,
\, w_i$ and $q_{ij}$) into Eqs.(\ref{3.3}) and (\ref{3.4}), we
have
\begin{eqnarray}
&& 0 = c^2  - 2w + \frac{2 w^2 }{c^2}
  + \frac{8w_i}{c^2}\frac{dx^i}{dt}
  - \left[ 1 + \frac{2w}{c^2} + \frac{2w^2}{c^4} \right]
  \left| \frac{d{\bf x}}{dt} \right|^2
  - \frac{q_{ij}}{c^4}\frac{dx^i}{dt}\frac{dx^j}{dt}
  \, , \label{3.9} \\
&& \frac{d^2 x^i}{dt^2} = w_{,i} \left( 1 - \frac{4w}{c^2}
  +\frac{1}{c^2} \left| \frac{d{\bf x}}{dt} \right|^2 \right)
  \nonumber \\
&& \qquad \quad - 2\frac{dx^i}{dt}\left( \frac{3w_{,t}}{2c^2}
  + \frac{2w_{,j}}{c^2} \frac{dx^j}{dt}
  - \frac{2w_{(j,k)}}{c^4}\frac{dx^j}{dt}\frac{dx^k}{dt}
  - \frac{w_{,t}}{2c^4}\left| \frac{d{\bf x}}{dt} \right|^2
  \right) \nonumber \\
&& \qquad \quad  + \frac{4w_{i,t}}{c^2}
  + \frac{8w_{[i,j]}}{c^2}\frac{dx^j}{dt}
  - \frac{q_{ij,k}}{c^4}\frac{dx^j}{dt}\frac{dx^k}{dt}
  - \frac{q_{jk,i}}{2c^4}\frac{dx^j}{dt}\frac{dx^k}{dt}
  \, . \label{3.10}
\end{eqnarray}
We assume that the solution is ${\bf x} = {\bf x}_N + {\bf x}_{1P}
+ {\bf x}_{2P}$, where ${\bf x}_N $ is Newtonian solution
(straight line), ${\bf x}_{1P}$ and ${\bf x}_{2P}$ are 1PN and 2PN
post Newtonian revision respectively. Then we have a series of
relations:
\begin{eqnarray}
&& x^i = x^i_0 + cn^i (t - t_0 ) + x_{1p}^i
  + x_{2p}^i  \, ,\label{3.11} \\
&& \frac{dx^i}{{dt}} = cn^i +  \frac{dx_{1p}^i }{dt}
  +  \frac{dx_{2p}^i}{dt} \, ,\label{3.12} \\
&& \frac{d^2 x^i}{dt^2} = \frac{d^2 x_{1p}^i}{dt^2}
  + \frac{d^2 x_{2p}^i}{dt^2} \, ,\label{3.13} \\
&& \left| \frac{d{\bf x}}{dt} \right|^2  = c^2
  + 2c {\bf n} \cdot \frac{d{\bf x}_{1p}}{dt}
  + 2c {\bf n} \cdot \frac{d{\bf x}_{2p}}{dt}
  + \left| \frac{d{\bf x}_{1p}}{dt} \right|^2
  \, ,\label{3.14} \\
&& \frac{dx^i}{dt}\frac{dx^j}{dt} = c^2 n^i n^j
  + c n^i \frac{dx_{1p}^j}{dt}
  + c n^j \frac{dx_{1p}^i}{dt}
  + c n^i \frac{dx_{2p}^j}{dt}
  + c n^j \frac{dx_{2p}^i} {dt}
  +  \frac{dx_{1p}^i}{dt}\frac{dx_{1p}^j}{dt}
  \, ,\label{3.15}
\end{eqnarray}
where we have neglected all of the terms higher than 2PN in
Eqs.(\ref{3.14}) and (\ref{3.15}). Substituting Eqs.(\ref{3.12}),
(\ref{3.14}) and (\ref{3.15}) into Eq.(\ref{3.9}), and considering
Eq.(\ref{3.8a}), we obtain
\begin{equation}\label{3.16}
{\bf n} \cdot \frac{d{\bf x}_{2P}}{dt}
  = \frac{4w_i}{c^2}n^i
  - \frac{1}{2c}\left| \frac{d{\bf x}_{1P}}{dt} \right|^2
  + \frac{4w^2}{c^3} - \frac{q_{ij}}{2c^3}n^i n^j \, ,
\end{equation}
where $q_{ij}= q_{\langle ij \rangle } + \delta _{ij} q/3$, and
$q_{\langle ij \rangle }$ is a trace-free symmetric tensor. In the
global coordinates of the solar system, $q_{\langle ij\rangle}=0$,
therefore $q_{ij}=\delta_{ij}q/3$. In Eq.(\ref{3.16}) ${\bf
x}_{1P}$ can be solved from Eqs.(\ref{3.8a}) and (\ref{3.8b}),
then Eq.(\ref{3.16}) becomes
\begin{equation}\label{3.17}
{\bf n} \cdot \frac{d{\bf x}_{2P}}{dt}
  = \frac{4w_i}{c^2}n^i
  - \frac{1}{2c} \left| \frac{d{\bf x}_{1P}}{dt} \right|^2
  + \frac{4w^2}{c^3} - \frac{q}{6c^3} \, .
\end{equation}
Substituting Eqs.(\ref{3.8a}), (\ref{3.12}), (\ref{3.13}),
(\ref{3.14}) and (\ref{3.15}) into Eq.(\ref{3.10}), we have
\begin{eqnarray}
\frac{d^2 x_{2P}^i}{dt^2} &=& -\frac{2}{c}w_{,t}n^i
  + \frac{4}{c}w_{(j,k)} n^in^jn^k
  + \frac{8}{c}w_{[j,k]} n^j - \frac{8}{c^2}ww_{,i}
  - \frac{4}{c} \left( \frac{d{\bf x}_{1P}}{dt}
   \cdot \nabla w \right) n^i \nonumber \\
 &&  - \frac{4}{c} \left( {\bf{n}}
  \cdot \nabla w \right) \frac{dx_{1P}^i}{dt}
  -\frac{1}{c^2} q_{ij,k} n^j n^k
  + \frac{1}{2c^2 }q_{jk,i} n^j n^k
  \, . \label{3.18}
\end{eqnarray}
If $q_{ij}=\delta _{ij}q/3$ is taken, we obtain
\begin{eqnarray}
\frac{d^2 x_{2P}^i}{dt^2} &=& -\frac{2}{c}w_{,t}n^i
  + \frac{4}{c}w_{(j,k)} n^in^jn^k
  + \frac{8}{c}w_{[j,k]} n^j - \frac{8}{c^2}ww_{,i}
  - \frac{4}{c} \left( \frac{d{\bf x}_{1P}}{dt}
   \cdot \nabla w \right) n^i \nonumber \\
 &&  - \frac{4}{c} \left( {\bf{n}}
  \cdot \nabla w \right) \frac{dx_{1P}^i}{dt}
  -\frac{1}{3c^2}\left( {\bf n}\cdot\nabla q \right) n^i
  + \frac{1}{6c^2 }q_{,i}
  \, . \label{3.19}
\end{eqnarray}
Eqs.(\ref{3.16}) and (\ref{3.18}) are just the 2PN equations of
light ray that we expected. By means of Eqs.(\ref{3.16}) and
(\ref{3.18}), based on Eqs.(\ref{3.8a}) and (\ref{3.8b}), we can
discuss light propagation at the 2PN level by means of the
iterative method. In the special case of the solar system, we may
use a much simple formulae (Eqs.(\ref{3.17}) and (\ref{3.19})),
where $Q$ (or $q$) is presented in Eq.(\ref{61eq}) \cite{gong04}.

The 1PN term ${\bf x}_{1P}$ in the equations can be obtained from
the solutions of the 1PN equations of light ray, i.e. from
Eqs.(\ref{3.8a}) and (\ref{3.8b}). Then the 2PN equations are
solvable. Furthermore, if we knew ever higher order metric tensor
(higher than 2PN), we also could get higher order equations of
light ray by means of such iterative method. Besides, the
iterative method can also be discussed on the parametrized 2PN
equations of light ray, then we can discuss the 2PN equations of
light ray in alternative gravitational theories which we will
discuss in the following section.


\section{Parametrized 2PN Equations of Light Ray}

The extension may also be used in parametrized post Newtonian
(PPN) formalism, which would be found a widely use for the
relativistic experiments and observation \cite{will93}. In terms
of more exact consideration \cite{klio00}, PPN extension for DSX
scheme (1PN) is written as
\begin{eqnarray}
g_{00} &=& -1 + \frac{2w}{c^2} - \frac{2}{c^4} \beta w^2
  + O(6) \, , \label{4.1a} \\
g_{0i} &=&  - \frac{2(1 + \gamma )}{c^3} w^i
  + O(5) \, , \label{4.1b} \\
g_{ij} &=& \delta _{ij} \left( 1 + \frac{2\gamma }{c^2} w \right)
  + O(4) \, . \label{4.1c}
\end{eqnarray}
Comparing to a known parametrized second post Newtonian formalism
\cite{euba99} our extension for parametrized second post Newtonian
 formalism is taking a tentative form in the extended 2PN DSX scheme
\begin{eqnarray}
g_{00} &=& -1 + \frac{2w}{c^2} - \frac{2}{c^4} \beta w^2
  + O(6) \, , \label{4.2a}\\
g_{0i} &=&  - \frac{2(1 + \gamma )}{c^3} w^i
  + O(5) \, , \label{4.2b}\\
g_{ij} &=& \delta _{ij} \left( 1 + \frac{2\gamma }{c^2}w \right)
  + \frac{\delta_{ij} \varepsilon}{c^4}\left( 2w^2+\frac{q}{3} \right)
  + \frac{\eta}{c^4}q_{ < ij > }  + O(6)  \, ,\label{4.2c}
\end{eqnarray}
When $\gamma = \beta = \varepsilon = \eta = 1$, then the metric
tensor return to the case in general relativity. If we ignore all
of $1 / c^4$ terms in $g_{ij}$, our results agree with the one of
Ref.\cite{klio00}

From the parametrized 2PN metric tensor we can get its
contravariant tensor as following
\begin{eqnarray}
g^{00} &=&  - 1 - \frac{2w}{c^2} + \frac{2(\beta - 2)w ^2 }{c^4}
  + O(6) \, , \label{4.3a} \\
g^{0i} &=& - \frac{2(1 + \gamma )}{c^3}w_i + O(5)\, , \label{4.3b} \\
g^{ij} &=& \delta _{ij} \left( 1 - \frac{2\gamma}{c^2}w \right)
  + \frac{\delta _{ij}}{c^4}\left[ (4\gamma ^2 - 2\varepsilon )w ^2
  - \frac{\varepsilon q}{3} \right] - \frac{\eta}{c^4}q_{ < ij > }
  + O(6)\, . \label{4.3c}
\end{eqnarray}
Then, the Parametrized 2PN Christoffel symbols:
\begin{eqnarray}
\Gamma _{00}^0 &=& - \frac{w _{,t}}{c^3} + O(5) \, ,
  \label{4.4a} \\
\Gamma _{0i}^0 &=& -\frac{w _{,i}}{c^2}
  +\frac{2(\beta -1)w w_{,i}}{c^4} + O(6) \, , \label{4.4b} \\
\Gamma _{00}^i &=& - \frac{w _{,i}}{c^2}
  - \frac{2(1 + \gamma )w_{i,t}}{c^4}
  + \frac{2(\beta + \gamma )w w _{,i}}{c^4} + O(6)
  \, , \label{4.4c} \\
\Gamma _{ij}^0 &=& \frac{2(1 + \gamma )}{c^3}w _{(i,j)}
  + \delta_{ij} \frac{\gamma w _{,t}}{c^3} + O(5)
  \, , \label{4.4d} \\
\Gamma _{0j}^i &=& - \frac{2(1 + \gamma )}{c^3}w _{[i,j]}
  + \delta _{ij} \frac{\gamma w _{,t}}{c^3} + O(5)
  \, , \label{4.4e} \\
\Gamma _{jk}^i &=& (\frac{\gamma }{c^2}
  + \frac{2(\varepsilon - \gamma^2) w}{c^4})(\delta_{ij} w_{,k}
  + \delta _{ik} w_{,j} - \delta _{jk} w_{,i} )
  + \frac{\varepsilon }{6c^4} \left( \delta _{ij} q_{,k}
  + \delta _{ik}q_{,j}  - \delta _{jk} q_{,i} \right)
  \nonumber \\
  && + \frac{\eta }{2c^4}(q_{ < ij > ,k}
  + q_{ < ik > ,j} - q_{ < jk > ,i} ) + O(6)
  \, . \label{4.4f}
\end{eqnarray}
Substituting the 1PN part of Eqs.(\ref{4.2a})-(\ref{4.4f}) into
Eqs.(\ref{3.3}) and (\ref{3.4}), we get the parametrized 1PN
equations of light ray:
\begin{eqnarray}
&& c{\bf n} \cdot \frac{d{\bf x}_{1P}}{dt} = - (1 + \gamma )w
  \, , \label{4.5} \\
&& \frac{d^2 {\bf x}_{1P}}{dt^2} = (1 + \gamma )\left[
  \nabla w - 2{\bf n}({\bf n} \cdot \nabla w) \right]
  \, , \label{4.6}
\end{eqnarray}
which have been shown in Ref.\cite{will93} (see Eq.(6.14) and
(6.15) of \cite{will93}).

Considering ${\bf x} = {\bf x}_0 + {\bf n} (t-t_0) + {\bf x}_{1P}
+ {\bf x}_{2P}$, Eqs.(\ref{4.5}), (\ref{4.6}) and substituting
whole Eqs.(\ref{4.2a})-(\ref{4.4f}) into Eqs.(\ref{3.3}) and
(\ref{3.4}) we get the parametrized 2PN equations of light ray:
\begin{eqnarray}
&& {\bf n} \cdot \frac{d{\bf x}_{2P}}{dt}
  = \frac{2(1 + \gamma )w_i}{c^2}n^i
  - \frac{1}{2c}\left| \frac{d{\bf x}_{1P}}{dt} \right|^2
  + \frac{(\beta +2\gamma +2\gamma ^2 -\varepsilon )w^2}{c^3}
  - \frac{\varepsilon q}{6c^3}
  - \frac{\eta  q_{\langle ij\rangle }}{2c^3}n^i n^j
  \, , \label{4.7} \\
&& \frac{d^2 x_{2P}^i}{dt^2}=-\frac{(1 + \gamma )}{c}w_{,t}n^i
  + \frac{2(1 + \gamma )}{c}w_{(j,k)} n^in^jn^k
  + \frac{4(1 + \gamma )}{c}w_{[i,j]} n^j \nonumber \\
  && \qquad \qquad - \frac{2(\beta + 2\gamma
  + 2\gamma^2 -\varepsilon)}{c^2}ww_{,i}
  - \frac{2(1 + \gamma )}{c}(\frac{d{\bf x}_{1P}}{dt}
  \cdot \nabla w)n^i - \frac{2(1 + \gamma )}{c}({\bf{n}}
  \cdot \nabla w)\frac{dx_{1P}^i}{dt} \nonumber \\
  && \qquad \qquad - \frac{4(\varepsilon
  - \gamma ^2 - \beta  + 1)}{c^2 }
  w({\bf n} \cdot \nabla w)n^i
  - \frac{\eta }{c^2}q_{\langle ij\rangle ,k} n^j n^k
  -\frac{\varepsilon }{3c^2}q_{,k} n^i n^k \nonumber \\
  && \qquad \qquad + \frac{\eta }{2c^2}q_{\langle jk\rangle ,i}
  n^jn^k + \frac{\varepsilon }{{6c^2 }}q_{,i}
  \, . \label{4.8}
\end{eqnarray}
In special case of the solar system, we have $q_{\langle ij
\rangle } =0$, therefore the equations become
\begin{eqnarray}
&& {\bf n} \cdot \frac{d{\bf x }_{2P}}{dt}
  = \frac{2(1 + \gamma )w_i }{c^2 }n^i
  - \frac{1}{2c}\left| \frac{d{\bf x }_{1P}}{dt} \right|^2
  + \frac{(\beta +2\gamma +2\gamma^2 -\varepsilon )w^2}{c^3}
  - \frac{\varepsilon q}{6c^3}
  \, , \label{4.9} \\
&& \frac{d^2x_{2P}^i}{dt^2}=-\frac{(1 + \gamma )}{c}w_{,t} n^i
  + \frac{2(1 + \gamma )}{c}w_{(j,k)}n^in^jn^k
  + \frac{4(1 + \gamma )}{c}w_{[i,j]} n^j \nonumber \\
  && \qquad \qquad -\frac{2(\beta +2\gamma
  + 2\gamma^2 -\varepsilon)}{c^2} ww_{,i}
  - \frac{2(1 + \gamma )}{c}(\frac{d{\bf x}_{1P}}{dt}
  \cdot \nabla w)n^i - \frac{2(1 + \gamma )}{c}({\bf n} \cdot
  \nabla w)\frac{dx_{1P}^i}{dt} \nonumber \\
  && \qquad \qquad - \frac{4(\varepsilon - \gamma ^2
  - \beta + 1)}{c^2} w({\bf n} \cdot \nabla w)n^i
  - \frac{\varepsilon }{3c^2} ({\bf n} \cdot \nabla q)n^i
  + \frac{\varepsilon }{6c^2}q_{,i} \, . \label{4.10}
\end{eqnarray}
When $\beta =\gamma =\varepsilon =\eta =1 $, Eqs.(\ref{4.7}),
(\ref{4.8}), (\ref{4.9}) and (\ref{4.10}) return to
Eqs.(\ref{3.16}), (\ref{3.18}), (\ref{3.17}) and (\ref{3.19})
respectively. Maybe in our parametrized 2PN equations of light
ray, four parameters are not sufficient. But if we compare with
Ref.\cite{euba99}, these four parameters are main parameters in
the 2PN equations of light ray.


\section{Conclusion Remarks}

1. We have extended DSX scheme to 2PN level approximation, by
means of which we can discuss the 2PN contribution on the
trajectory of light ray. Our extension keep the main advantage in
DSX scheme (the linear equations of the potential and vector
potential, and the linear relationship between $w$, $w_i$ and $W$,
$W_a$). In the case of the solar system, we obtain a special
solution under the PN gauge. But we should point out that if we
want to discuss the 2PN contribution on the motion of bodies (not
light ray), we have to extend the metric further more ($q_{00}
\sim O(8)$ and $q_{0i} \sim O(7)$) which someone might finish in
future. Considering the precision of ASTROD, they will need the
theoretical precision up to the 2PN level for the motion of
bodies.

2. We have discussed 2PN equations of light ray based on the
extended DSX scheme by means of the iterative method. If we have
the higher order metric, we also can obtain the higher order
equations of light ray with a similar iterative method in this
paper. Therefore in some meaning the way shown in the section 3 is
general. The special case for equations of light ray propagating
in the solar system($q_{ij}= \frac{1}{3}q$) is also considered.

3. We also use a parametrized 2PN metric tensor (with four
parameters) and its corresponding Christoffel symbols to
substitute into the null geodetic equations and obtain
parametrized 2PN equations of light ray. Although we have not
induced four parameters strictly, but it agree with the main
parameter in \cite{euba99}. Also if 2PN part of $g_{ij}$ is
neglected, the metric tensor return to the one in \cite{klio00}
which has been deduced strictly.

Anyway, we hope our work to be useful for the measurement of
future space missions.


\vspace{0.6cm}
\noindent{\bf Acknowledgments}

\bigskip\noindent
This work was supported by the National Natural Science Foundation
of China (Grant No. 10273008). In the workshop ``Relativistic
Astrophysics and High-precision Astrodynamics" led by Prof.
Wei-tou Ni, we had several very fruitful discussions.



\end{document}